\documentclass[aps,prl,reprint,showpacs,floatfix,citeautoscript,longbibliography,superscriptaddress]{revtex4-1}
\usepackage{graphicx}
\usepackage{bm,amsmath,amssymb,mathrsfs,dcolumn}
\usepackage[colorlinks=true, linkcolor=blue, citecolor=blue, urlcolor=blue, linktoc=page, bookmarks=false, pdfstartview={FitH}, pdfborder={0 0 0.0 [3 3]}]{hyperref} 
\usepackage[usenames]{xcolor} 
\hypersetup{pdfborder=0 0 0,colorlinks=true,citecolor=blue,linkcolor=blue}
\usepackage[final]{pdfpages}

\begin{document}
\title{Giant room temperature interface spin Hall and inverse spin Hall effects}
\author{Lei Wang}
\altaffiliation{These authors contributed equally to this work.}
\affiliation{The Center for Advanced Quantum Studies and Department of Physics, Beijing Normal University, 100875 Beijing, China}
\affiliation{Faculty of Science and Technology and MESA$^+$ Institute for Nanotechnology, University of Twente, P.O. Box 217, 7500 AE Enschede, The Netherlands}
\author{R. J. H. Wesselink}
\altaffiliation{These authors contributed equally to this work.}
\affiliation{Faculty of Science and Technology and MESA$^+$ Institute for Nanotechnology, University of Twente, P.O. Box 217, 7500 AE Enschede, The Netherlands}
\author{Yi Liu}
\affiliation{The Center for Advanced Quantum Studies and Department of Physics, Beijing Normal University, 100875 Beijing, China}
\affiliation{Faculty of Science and Technology and MESA$^+$ Institute for Nanotechnology, University of Twente, P.O. Box 217, 7500 AE Enschede, The Netherlands}
\author{Zhe Yuan}
\email[To whom correspondence should be addressed: ]{zyuan@bnu.edu.cn}
\affiliation{The Center for Advanced Quantum Studies and Department of Physics, Beijing Normal University, 100875 Beijing, China}
\author{Ke Xia}
\affiliation{The Center for Advanced Quantum Studies and Department of Physics, Beijing Normal University, 100875 Beijing, China}
\author{Paul J. Kelly}
\affiliation{Faculty of Science and Technology and MESA$^+$ Institute for Nanotechnology, University of Twente, P.O. Box 217, 7500 AE Enschede, The Netherlands}
\date{\today}
\begin{abstract}
The spin Hall angle (SHA) is a measure of the efficiency with which a transverse spin current is generated from a charge current by the spin-orbit coupling and disorder in the spin Hall effect (SHE).  In a study of the SHE for a Pt$|$Py (Py=Ni$_{80}$Fe$_{20}$) bilayer using a first-principles scattering approach, we find a SHA that increases monotonically with temperature and is proportional to the resistivity for bulk Pt. By decomposing the room temperature SHE and inverse SHE currents into bulk and interface terms, we discover a giant interface SHA that dominates the total inverse SHE current with potentially major consequences for applications.
\end{abstract}
\pacs{72.25.Ba,  
          72.25.Mk, 
          75.70.Tj,   
          85.75.-d    
         }
\maketitle

{\it\color{red}Introduction.---}The spin Hall effect (SHE) refers to the generation of a transverse spin current by an electrical current flowing in a conductor \cite{Dyakonov:pla71, Hirsch:prl99, Vignale:jsnm10}. It and its inverse (the ISHE) allow pure spin currents to be created, manipulated and detected electrically \cite{Hoffmann:ieeem13, Sinova:rmp15, Niimi:rpp15} and are extensively applied in spintronics experiments and devices \cite{Uchida:natm10, Jungwirth:natm12, Liu:sc12, Nakayama:prl13, Zhang:natp15}. The SHE is a relativistic effect that results from disorder scattering by impurities \cite{Engel:prl05}, phonons \cite{Gorini:prl15}, and surface roughness \cite{Zhou:prb15} in combination with spin-orbit coupling. Quantitative studies have focused on disentangling ``intrinsic'' and ``extrinsic'' mechanisms at low temperatures in bulk metals and alloys \cite{Guo:prl08, Tanaka:prb08, Freimuth:prl10, Lowitzer:prl11, Zimmermann:prb14}, where ``intrinsic'' refers to scattering that can be related to the electronic structure of the perfectly ordered material while ``extrinsic'' refers to disorder induced scattering \cite{Vignale:jsnm10,Sinova:rmp15}. The intrinsic spin Hall conductivity of bulk Pt is predicted to decrease with increasing temperature \cite{Guo:prl08}. Phenomenological theories predict different dependence on the mobility for extrinsic side-jump and skew-scattering mechanisms \cite{Engel:prl05, Gorini:prl15, Sinova:rmp15}. However, it was recently demonstrated that phenomenological theories derived at zero temperature can not be directly extrapolated to finite temperatures \cite{Gorini:prl15}. Separating the intrinsic and extrinsic contributions in experiment turns out to be very difficult \cite{Vila:prl07, Isasa:prb15}. 

The efficiency of converting charge current to spin current is expressed in terms of the spin current per unit charge current, the spin-Hall angle (SHA). For the important and much studied heavy element Pt there is an unsatisfactorily large spread in the magnitude of the values reported for the SHA \cite{Hoffmann:ieeem13, Sinova:rmp15, Niimi:rpp15}. Even though all experiments involve an interface between a ferromagnetic (FM) material and a nonmagnetic (NM) metal \cite{Valenzuela:nat06, Saitoh:apl06, Vila:prl07, Kimura:prl07, Liu:prl11, Ando:jap11, Mosendz:prb10, Kim:prb14,Isasa:prb15,Pai:prb15}, this is not considered in the theoretical models used to extract the SHA. Very recent experiments \cite{Rojas-Sanchez:prl14} and theory \cite{Liu:prl14} however show that neglecting interface spin memory loss leads to an underestimation of the bulk spin-flip diffusion length. In this Letter we explicitly include an interface in a study of the SHE. To do so we develop a first-principles computational scheme to calculate local longitudinal and transverse currents in a scattering geometry and apply it to the study of the SHE in pure ``bulk'' Pt and in a Py$|$Pt bilayer (Py=Ni$_{80}$Fe$_{20}$). The temperature-dependent SHA of bulk Pt is found to exhibit a linear dependence on the electrical resistivity. From the Py$|$Pt bilayer calculations we extract a value of the SHA for bulk Pt that is consistent with the pure, bulk value while the interface makes a significant contribution to both the SHE and ISHE that should be taken into account in interpreting experiments.

\begin{figure}[t]
\includegraphics[width=\columnwidth]{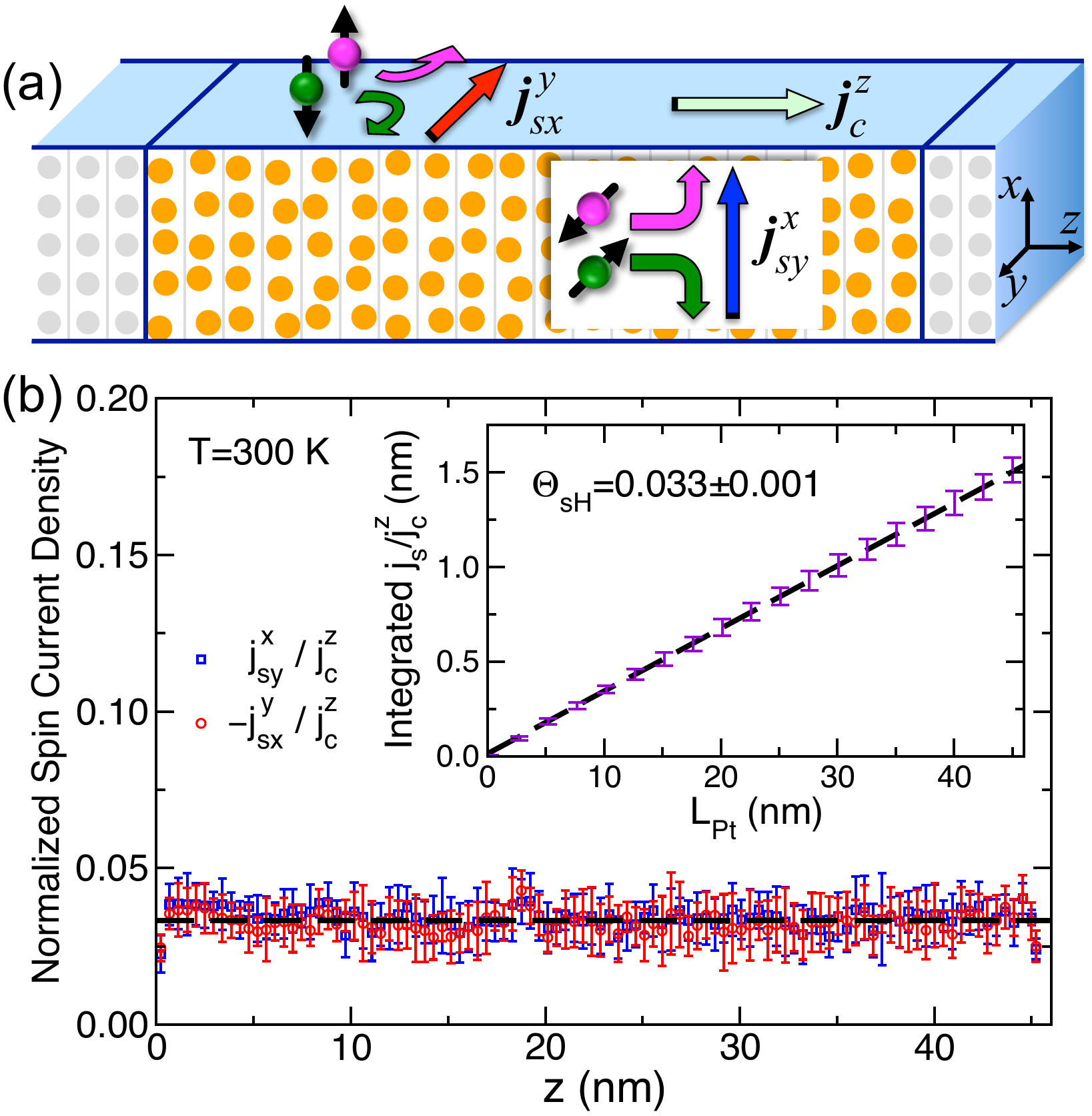}
\caption{(a) Schematic illustration of the scattering geometry. Electrons flow ($j_c^z$) from the perfectly crystalline left lead to the right one through a disordered region of pure bulk Pt where atoms are displaced from their equilibrium positions by populating phonons \cite{Liu:prb15}. Transverse spin currents arising from the SHE flowing along $x$ and $-y$ directions are polarized in the $y$ ($j^x_{sy}$) and $x$ ($j^y_{sx}$) directions, respectively. (b) Calculated transverse spin-current densities in Pt normalized by $j_c^z$ at room temperature. The error bars are a measure of the spread of 10 random configurations. The dashed black line shows the extracted SHA. Inset: integrated spin current density in a length $L_{\rm Pt}$ of disordered Pt. The dashed black line illustrates a linear least squares fit from which a SHA for pure bulk Pt of $\Theta_{\rm sH}=0.033\pm0.001$ is extracted.}\label{fig:1}
\end{figure}

{\it\color{red}SHE for Pt.---}We set up a scattering geometry 
\footnote{In the linear response regime of interest here, the scattering formalism we are using [\onlinecite{Khomyakov:prb05}] is equivalent to the Kubo formalism.} 
consisting of two crystalline semi-infinite Pt leads sandwiching a scattering region of disordered Pt with atoms displaced from their equilibrium positions by populating phonon modes \cite{Liu:prb15}, as sketched in Fig.~\ref{fig:1}(a). For the resistivity and spin-flip diffusion length, this approach has been shown to yield essentially perfect agreement with experiment \cite{Liu:prb15}. To study the SHE, we calculate the local longitudinal and transverse charge and spin current densities in the scattering region so that both intrinsic and extrinsic contributions are naturally included.

Within the framework of density-functional theory, the electronic structure of bulk Pt is calculated using tight-binding linear muffin-tin orbitals (TB-LMTO) in combination with the atomic spheres approximation. Electron and spin current densities are obtained as expectation values of the velocity operator 
$\hat{\bf v}=[\hat{\mathcal R},\hat{\mathcal H}]/(i\hbar)$ \cite{Turek:prb02, Wang:prb08} and evaluated using the Hamiltonian matrix $\hat{\mathcal H}_{\bf R\bf R'}$ for real-space TB-LMTOs and the position operator 
$\hat{\mathcal R}_{\bf R\bf R'}=\bf R\delta_{\bf R\bf R'}$ where $\bf R$ is the real-space position of atoms. To determine the local electron current density $\mathbf j_c$ from atom $\bf R'$ to atom $\bf R$ ($\mathbf R'$$\ne$$\mathbf R$), we calculate the expectation value of the corresponding block of the velocity matrix $(\mathbf R-\mathbf R')(\hat{\mathcal H}_{\bf R\bf R'}-H.c.)$ over all scattering states $\Psi_i$ determined by the ``wave-function matching'' scheme \cite{Ando:prb91, Xia:prb06} to eventually arrive at
${\bf j}_c=2({\bf R-R'})\sum_i{\rm Im}\langle\Psi_i\vert\hat{\mathcal H}_{\bf R\bf R'}\vert\Psi_i\rangle/\hbar$. 
The spin current density is obtained by replacing $\hat{\bf v}$ by $(\hat{\bm \sigma}\otimes\hat{\bf v}+\hat{\bf v}\otimes\hat{\bm \sigma})/2$ \cite{Wang:prb08, Haney:prl10}. We consider transport along an fcc [111] direction ($z$) and use a $5\times5$ lateral supercell with periodic boundary conditions in the $x$ and $y$ directions. The supercell Brillouin zone (BZ) is sampled with $64\times64$ $k$ points. The disordered Pt scattering region is $\sim 45$nm long and contains 200 atomic layers; ten random disorder configurations were calculated.

The longitudinal electron current density $j_c^z(z)$ along $z$ 
\footnote{We denote by $j_c^i$ the electron current density in the $i=x,y,z$ direction and by $j^i_{sk}$ the spin current density flowing along $i$ and polarized in the $k$ direction. All current densities are represented in units of particle current density, i.e. charge and spin (angular momentum) current densities are $-e j_c^i$ ($e>0$) and $j^i_{sk}\hbar/2$, respectively.} 
across an arbitrary plane is always a constant and equal to the total current density calculated from the scattering matrix. The local transverse current densities are projected onto the boundary of the lateral supercell with a cross section corresponding to an atomic layer. For a current of charge in the $z$ direction, the transverse spin currents in the $x$ and $-y$ directions should be polarized along $y$ and $x$, respectively, and have the same amplitudes. This expectation is borne out by the calculated local transverse spin current densities plotted in Fig.~\ref{fig:1}(b). The fluctuations arise from random configurations of phonon-induced atom displacements. 

To extract the SHA $\Theta_{\rm sH}$, we integrate the local spin current density (averaged over $x$ and $y$ directions) from the left boundary of the disordered Pt to a certain length $L_{\rm Pt}$. The result $\int_0^{L_{\rm Pt}}dz(j^x_{sy}-j^y_{sx})/(2j_c^z)$ is shown in the inset to Fig.~\ref{fig:1}(b). By analogy with the resistance $R=r_{\rm if}+\rho\cdot L$ of the lead$|$scattering region$|$lead geometry that is a sum of interface $r_{\rm if}$ and bulk $\rho\cdot L$ terms \cite{Starikov:prl10, Liu:prb11}, this integral contains both interface and bulk contributions. For sufficiently large $L_{\rm Pt}$, the integral is proportional to $L_{\rm Pt}$ and the slope corresponds to the SHA of pure bulk Pt, $\Theta_{\rm sH}=0.033\pm0.001$. Extraction in this way reduces the effect of fluctuations of the local transverse spin currents and a possible contribution from the interfaces between disordered Pt and the crystalline Pt leads. The latter is seen to be very small; only the first layers at either end show slightly lower transverse spin currents and the intercept of the fitting curve is essentially zero. 

\begin{figure}[b]
\includegraphics[width=\columnwidth]{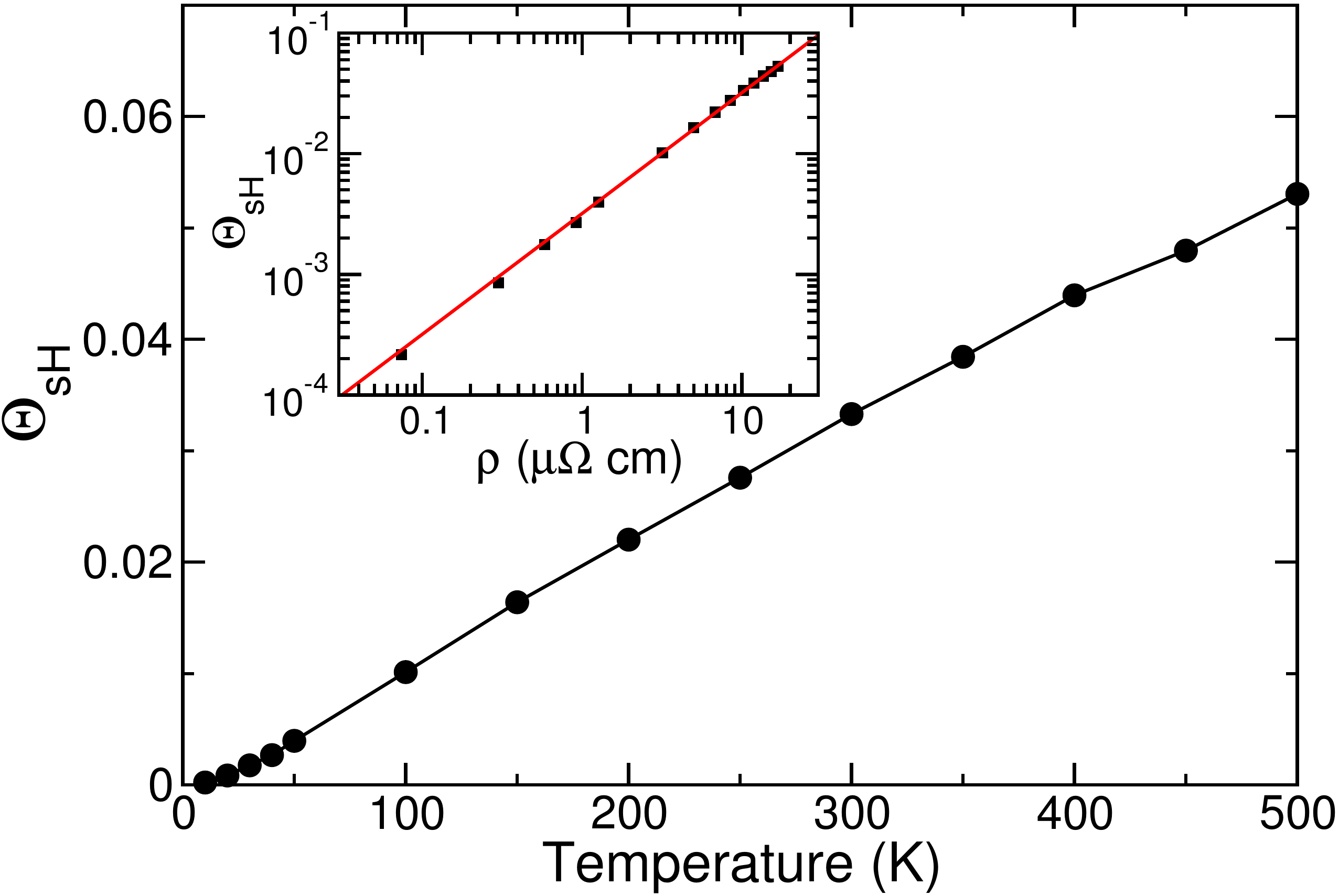}
\caption{Calculated SHA of pure bulk Pt as a function of temperature. Inset: calculated SHA replotted as a function of electrical resistivity on a log-log scale. The solid red line illustrates a linear dependence.}\label{fig:2}
\end{figure}
$\Theta_{\rm sH}$ is seen in Fig.~\ref{fig:2} to increase monotonically with temperature in qualitative agreement with the phenomenological theory of phonon skew scattering \cite{Gorini:prl15}. In the low-temperature limit very little scattering occurs, the system becomes ballistic, and all the Bloch states in the ballistic system propagate with fixed momenta and there is no transverse spin current \cite{Vignale:jsnm10}. We replot $\Theta_{\rm sH}$ as a function of electrical resistivity in the inset to Fig.~\ref{fig:2} where a perfectly linear dependence is seen from very low temperature up to 500~K. This proportionality indicates that the spin-Hall conductivity of Pt is nearly a constant, $\sigma_{\rm sH}=3.2\times 10^{5}(\hbar/2e)(\Omega\,\mathrm{m})^{-1}$. In previous work \cite{Liu:prl14, Liu:prb15}, we demonstrated that the spin-flip diffusion length $l_{\rm sf}$ of Pt is proportional to the conductivity. Thus the product of SHA and spin-flip diffusion length is nearly constant, $\Theta_{\rm sH}\cdot l_{\rm sf}\approx0.2$~nm. At this point it is also interesting to note that recently reported giant SHAs were found in $\beta$ phases of Ta \cite{Liu:sc12} and W \cite{Pai:apl12} which have complex unit cells and are highly resistive.

{\it\color{red}SHE and ISHE in a Py$|$Pt bilayer.---}To exploit the spin current  generated by the SHE in a NM metal like Pt, a FM material is usually attached to the NM metal so the spin current can be injected through the interface into the FM material to excite or switch the magnetization \cite{Liu:sc12, Zhang:natp15, Pai:prb15}. Such FM$|$NM bilayers are also widely used in spin-pumping experiments where a spin current is pumped by forced precession of the magnetization of the FM layer into the NM metal where it is detected by measuring the inverse spin-Hall voltage \cite{Saitoh:apl06, Mosendz:prb10, Ando:jap11, Rojas-Sanchez:prl14}. To explicitly investigate the effect of the FM$|$NM interface on the SHE and ISHE in the NM metal, we calculate the SHE and ISHE simultaneously in a Py$|$Pt bilayer. Specifically, we compute (i) the transverse spin currents in Pt that result from a longitudinal charge current (SHE) and (ii) the transverse charge current in Pt induced by the spin-polarized current injected from Py into Pt (ISHE).

We construct a bilayer consisting of 30 nm thick Pt and 10 nm thick Py that are modeled with room temperature \cite{Liu:prb15} lattice (Pt, Py) and spin (Py) disorder \footnote{Omitting the spin disorder in Py leads to more than a 20\% decrease in the extracted interface SHA.}. The interface is along the fcc (111) direction, we drive a charge current in the $z$ direction through the interface and study the response. In our lateral supercell, we use 9$\times$9 interface unit cells of Py to match $3\sqrt{7}\times3\sqrt{7}$ unit cells of Pt. The bilayer is connected to semi-infinite Cu electrodes with the same lattice constant as Py. The data we will show result from averaging 30 random configurations of disorder. $42\times42$ $k$ points were used to sample the supercell 2D BZ, the reciprocal space density being comparable to that used for pure Pt.

\begin{figure}[t]
\includegraphics[width=\columnwidth]{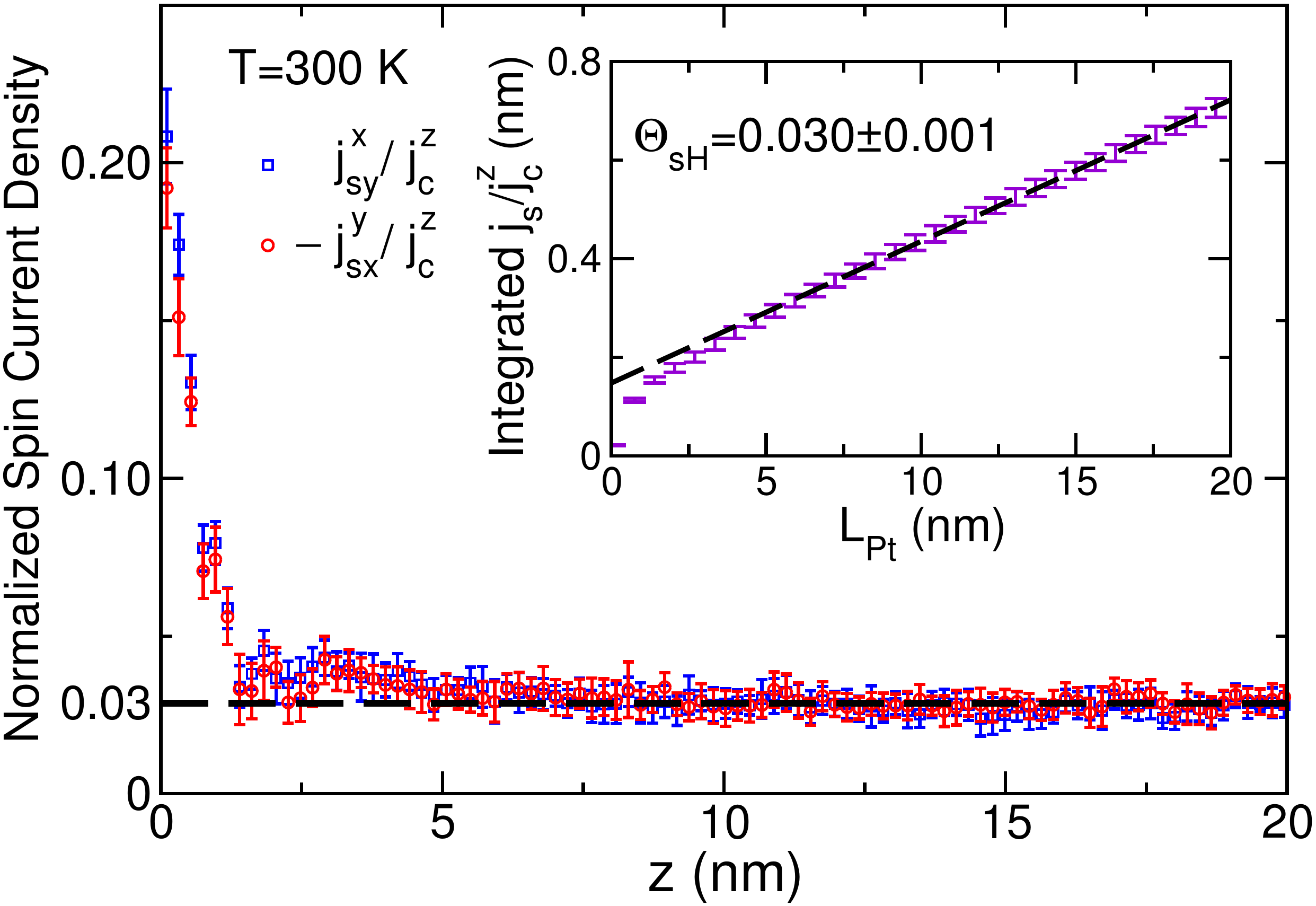}
\caption{SHE in a Py$|$Pt bilayer at room temperature. The calculated transverse spin current densities ($j_{sx}^y$ and $j^x_{sy}$) normalized by the longitudinal electron current density are shown as a function of the distance from the interface in the Pt part of a Py$|$Pt bilayer. Near the interface at $z=0$, the transverse spin current densities are much larger than the asymptotic bulk Pt values. The error bars measure the spread found for 30 random disorder configurations. The dashed black line shows the value of $\Theta_{\rm sH}$ extracted for bulk Pt. Inset: integrated spin current density in Pt, where the bulk SHA $\Theta_{\rm sH}=0.030\pm0.001$ is determined from the slope of a linear least squares fit shown by the dashed line.
}\label{fig:3}
\end{figure}

Figure~\ref{fig:3} shows the calculated transverse spin current densities $j^x_{sy}$ and $j^y_{sx}$ in Pt normalized by the longitudinal electron current density $j_c^z$. Sufficiently far from the interface, the transverse spin current density is nearly constant corresponding to the bulk SHA. To extract a value for $\Theta_{\rm sH}$, we integrate the transverse spin current densities from the Py$|$Pt interface to a length $L_{\rm Pt}$ into Pt and plot the result in the inset to Fig.~\ref{fig:3}. A linear least squares fit yields a bulk SHA, $\Theta_{\rm sH}=0.030\pm0.001$, in good agreement with the calculation for pure Pt. The small difference can be attributed to stretching Pt slightly to match to Py \cite{Liu:prl14}. The large interface contribution leads to a finite intercept in the linear fit of the integrated spin current shown in the inset to Fig.~\ref{fig:3}.

\begin{figure}[t]
\includegraphics[width=\columnwidth]{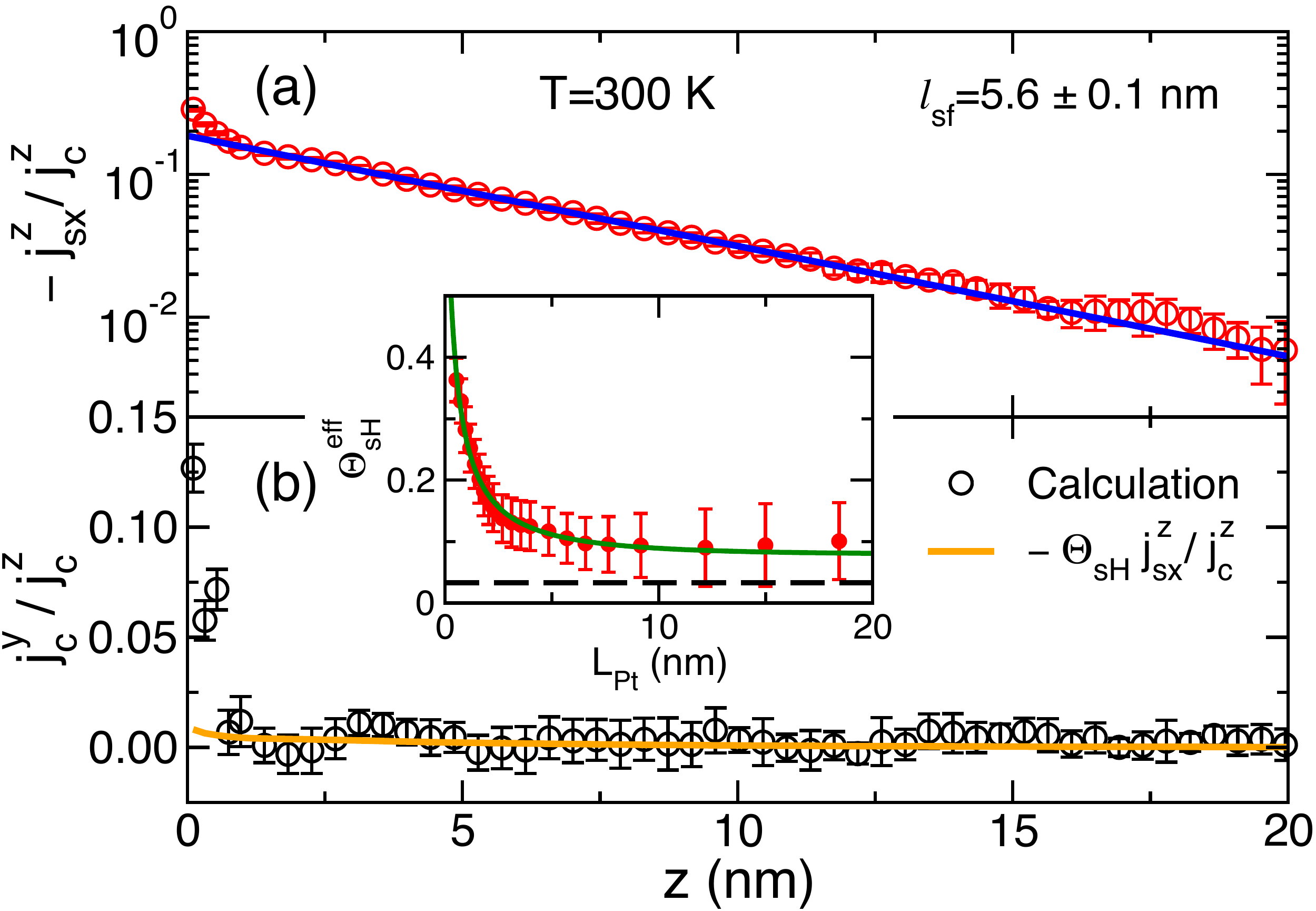}
\caption{ISHE in a Py$|$Pt bilayer at room temperature. (a) Spin current density injected from Py into Pt. The current is polarized along the $-x$ direction in Py and loses this polarization in Pt by spin-flip scattering. The Pt spin-flip diffusion length, $l_{\rm sf}=5.6\pm0.1$ nm, is obtained by an exponential fit (solid blue line). Note the logarithmic $y$ axis. (b) Transverse charge current density generated in Pt by the ISHE. The thick orange line shows the calculated longitudinal spin-current density multiplied by the extracted value of the bulk SHA ($\Theta_{\rm sH}=0.030$). Inset: effective SHA (red dots) calculated by integrating the corresponding transverse charge current density and longitudinal spin current density in a certain range of Pt ($0\le z\le L_{\rm Pt}$). The solid green line is a fit using Eq.~(\ref{eq:eff}).}
\label{fig:4}
\end{figure}

The longitudinal charge current injected into Pt is spin-polarized along the $-x$ axis by Py. The spin polarization, shown in Fig.~\ref{fig:4}(a), is described by the extended Valet-Fert theory \cite{Valet:prb93, Bass:jpcm07} according to which the polarization decreases exponentially as the distance from the interface because of spin-flip scattering. The spin-flip diffusion length $l_{\rm sf}=5.6\pm0.1$ nm obtained by an exponential fit (solid blue line) is in agreement with the value (5.5$\pm$0.2 nm) obtained by injecting a perfectly spin polarized current into bulk Pt at room temperature \cite{Liu:prl14}. At the interface, the spin polarization shows a much faster decay corresponding to the spin memory loss at the Py$|$Pt interface \cite{Nguyen:jmmm14, Rojas-Sanchez:prl14, Liu:prl14}.

The spin current along the $z$ axis leads by virtue of the ISHE to a charge current in the $y$ direction. This transverse charge current $j_c^y$ is plotted in Fig.~\ref{fig:4}(b) (open black circles). In the bulk region, $j_c^y$ is consistent with the product (solid orange line) of the bulk value of $\Theta_{\rm sH}$ and the calculated longitudinal spin current $-j^z_{sx}$. At the interface, however, the bulk SHA results in a significant underestimate of the transverse charge current. Both the SHE shown in Fig.~\ref{fig:3} and the ISHE in Fig.~\ref{fig:4}(b) suggest a larger local SHA at the interface that we call $\Theta^I_{\rm sH}$.

{\it\color{red}Interface SHA.---}In an ISHE experiment, the total charge current (voltage) is measured including both interface and bulk contributions which, in general, can be different. Indeed, an experiment has been reported in the literature where the Py$|$Bi interface shows an opposite SHA to that of bulk Bi \cite{Hou:apl12}. In a recent first-principles calculation for a Co$|$Pt bilayer, it was found that the SHA of each atomic Pt layer depends on the distance to the Co$|$Pt interface and the SHA of the interface Pt layer is much larger than that of the interior Pt layers \cite{Freimuth:prb15}. To interpret the ISHE for a FM$|$NM bilayer and obtain a quantitative value of the interface SHA $\Theta^I_{\rm sH}$, we consider the segment of Pt extending from the interface at $z=0$ to a position $z=L_{\rm Pt}$ where the total transverse electron current $\bar J_c\equiv\int_0^{L_{\rm Pt}}j^y_c(z)dz$ is generated by the total spin current $\bar J_s\equiv\int_0^{L_{\rm Pt}}j^z_{sx}(z)dz$. We decompose these into bulk and interface contributions as follows. The longitudinal spin current can be represented as the  sum of an interface and bulk parts, $j_{sx}^z(z)=\bar J_s^I\delta(z)+j_s^0\exp(-z/l_{\rm sf})$. $j^0_s$ is defined as the value of the exponentially decaying spin current in Pt extrapolated back to the interface at $z=0$. $\bar J^I_s$ is the effective spin current density at the interface. Thus the total spin current density can be written as $\bar J_s=\bar J_s^I+j_{s}^0 l_{\rm sf}\left(1-e^{-L_{\rm Pt}/l_{\rm sf}}\right)$. The total transverse charge current density can be calculated with the interface and bulk SHAs as
\begin{eqnarray}
\bar J_c&=&\int_0^{L_{\rm Pt}}\left[\Theta^I_{\rm sH}\bar J_s^I\delta(z)+\Theta_{\rm sH}j_s^0 e^{-z/l_{\rm sf}}\right]dz\nonumber\\
&=&\Theta^I_{\rm sH}\bar J_{s}^I +\Theta_{\rm sH} j_{s}^0 l_{\rm sf}\left(1-e^{-L_{\rm Pt}/l_{\rm sf}}\right).
\end{eqnarray}
If we interpret the interface spin-Hall contribution in terms of an effective bulk value $\Theta^{\rm eff}_{\rm sH}\equiv\bar J_c/\bar J_s$, then
\begin{equation}
\Theta^{\rm eff}_{\rm sH}=
\frac{\Theta^I_{\rm sH} \bar J_s^I + \Theta_{\rm sH} 
       \left(1 - e^{-L_{\rm Pt}/l_{\rm sf}}\right) j_s^0 l_{\rm sf} }
{\bar  J_s^I + \left(1-e^{-L_{\rm Pt}/l_{\rm sf}} \right) j_s^0 l_{\rm sf} }.\label{eq:eff}
\end{equation}
$\Theta^{\rm eff}_{\rm sH}$ is plotted as a function of $L_{\rm Pt}$ in the inset to Fig.~\ref{fig:4} as red dots. Taking the bulk SHA, $\Theta_{\rm sH}=0.030$, and spin-flip diffusion length of Pt, $l_{\rm sf}=5.6$~nm, we are able to fit the calculated $\Theta^{\rm eff}_{\rm sH}$ using Eq.~(\ref{eq:eff}). The fit illustrated by the solid green line describes the calculated data points perfectly. The value we obtain for the interface SHA, $\Theta^I_{\rm sH}=0.87\pm0.07$, is some twenty five times larger than the bulk value. Even in the limit of thick Pt, $\Theta^{\rm eff}_{\rm sH}$ does not approach the bulk value $\Theta_{\rm sH}=0.030$ (dashed line) but saturates to a value of 0.08 indicating that it is essential to explicitly include an interface contribution.

The interface SHA also plays a crucial role in interpreting experiments where a spin current generated by the SHE is injected into an adjacent ferromagnet, e.g. using the spin-Hall torque to switch a magnetization \cite{Liu:sc12, Zhang:natp15, Pai:prb15}. The values of $\Theta^{\rm eff}_{\rm sH}$ extracted from these experiments are usually much larger than the bulk value $\Theta_{\rm sH}$ calculated in this work for two main reasons. The spin-flip diffusion length used to interpret the experiments is much smaller than the ``real'' value because of the neglect of interface spin memory loss \cite{Rojas-Sanchez:prl14, Liu:prl14}. The second reason is that the interface gives rise to a significant spin-Hall current that was attributed to the bulk. To increase the efficiency of converting charge current to spin current that is injected into an adjacent FM material for magnetization switching, our calculations suggest using highly resistive Pt (with substantial lattice disorder) (i) because of the correlation we find between the resistivity and $\Theta_{\rm sH}$ for bulk Pt in the inset to Fig.~\ref{fig:2} and (ii) because the charge current density may then be more concentrated at the interface for a better exploitation of the large interface SHA.

{\it\color{red}Conclusions.---}We have studied the SHE for pure bulk Pt and for a Py$|$Pt bilayer at finite temperature using a first-principles scattering formalism. The bulk SHA increases monotonically with increasing temperature and is proportional to the electrical resistivity. For a Py$|$Pt bilayer, the  (asymptotic) bulk SHA we extract agrees with that of pure Pt while the interface is found to play an important role for both the SHE and ISHE, a result suggested by some earlier studies \cite{Hou:apl12, Freimuth:prb15}. The interface SHA is some twenty five times larger than the bulk value. Reinterpretation of spin-Hall experiments with FM$|$NM bilayers in which interface contributions are taken into account properly is highly desirable. The insight provided by this work suggests exploiting the large SHA of FM$|$NM interfaces.

\begin{acknowledgments}
This work was triggered by a question asked by Frank Freimuth. We would like to thank helpful discussions with Gerrit Bauer and Guang-Yu Guo. This work was financially supported by National Basic Research Program of China (2011CB921803, 2012CB921304), NSF-China (11174037, 61376105), the Royal Netherlands Academy of Arts and Sciences (KNAW), the ``Nederlandse Organisatie voor Wetenschappelijk Onderzoek'' (NWO) through the research programme of ``Stichting voor Fundamenteel Onderzoek der Materie'' (FOM) and the  supercomputer facilities of NWO ``Exacte Wetenschappen (Physical Sciences)''. 
\end{acknowledgments}

\end{document}